\documentclass[aps,prl,preprint,onecolumn,superscriptaddress]{revtex4-1}
\usepackage{amsmath,amssymb}
\pdfoutput=1
\usepackage{graphicx}
\usepackage{float}
\usepackage{natbib}
\usepackage{color}
\usepackage[colorlinks=true,bookmarks=false,citecolor=blue,urlcolor=blue]{hyperref}
\usepackage{lmodern}
\usepackage[most]{tcolorbox}
\usepackage{tikz}
\usepackage{varwidth}
\usepackage{wrapfig}
\usepackage{capt-of}
\usepackage{paracol}
\usepackage{threeparttable}

\newcommand{\fp}{Fabry-P\'erot }
\newcommand{\JC}{Jaynes-Cummings }

\newcommand{\sq}{$^2$}

\begin{document}

\title{Casimir microcavities for tunable self-assembled polaritons}

\author{Battulga Munkhbat}
\affiliation{Department of Physics, Chalmers University of Technology, 412 96, Göteborg, Sweden}

\author{Adriana Canales}
\affiliation{Department of Physics, Chalmers University of Technology, 412 96, Göteborg, Sweden}

\author{Betül Küçüköz}
\affiliation{Department of Physics, Chalmers University of Technology, 412 96, Göteborg, Sweden}

\author{Denis G. Baranov}
\affiliation{Department of Physics, Chalmers University of Technology, 412 96, Göteborg, Sweden}

\author{Timur Shegai}
\email[]{timurs@chalmers.se}
\affiliation{Department of Physics, Chalmers University of Technology, 412 96, Göteborg, Sweden}

\begin{abstract}
\textbf{Hybrid light-matter states, polaritons, are one of the central concepts in modern quantum optics and condensed matter physics. Polaritons emerge as a result of strong interaction between an optical mode and a material resonance, which is frequently realized in molecular, van der Waals, or solid-state platforms \cite{lidzey1998strong,khitrova2006vacuum,torma2014strong,dai2014tunable,basov2016polaritons,sidler2017fermi,baranov2017novel}. However, this route requires accurate (nano)fabrication and often lacks simple means for tunability, which could be disadvantageous in some applications. Here, we use a different approach to realize polaritonic states by employing a stable equilibrium between two parallel gold nanoflakes in an aqueous solution \cite{chen2016rapid}. Such plates form a self-assembled \fp microcavity with the fundamental optical mode in the visible spectral range. The equilibrium distance between the plates is determined by a balance between attractive Casimir and repulsive electrostatic forces \cite{derjaguin1941acta,verwey1947theory,casimir1948attraction} and can be controlled by concentration of ligand molecules in the solution, temperature, and light pressure, which allows active and facile tuning of the cavity resonance by external stimuli. Using this Casimir approach, we demonstrate self-assembled polaritons by placing an excitonic medium in the microcavity region, as well as observe their laser-induced modulations in and out of the strong coupling regime. These Casimir microcavities can be used as sensitive and tunable polaritonic platforms for a variety of applications, including opto-mechanics \cite{eichenfield2009picogram}, nanomachinery \cite{zhao2019stable}, and cavity-induced effects, like polaritonic chemistry \cite{Angew16}}.
\end{abstract}

\maketitle

By studying the stability of colloidal solutions, Hendrik Casimir in 1948 came into conclusion about the existence of attractive forces between two parallel uncharged perfect electric conductor plates in vacuum, later named after him \cite{casimir1948attraction}. These forces are of quantum origin and exist even in the absence of any external charges or fields. They appear even at zero temperature and are a consequence of irremovable and unavoidable zero-point charge fluctuations \cite{rodriguez2015classical}. The related theory of colloidal stability was further developed into the so-called DLVO theory (after Derjaguin, Landau, Verwey and Overbeek), which main ingredients are attractive van der Waals and repulsive electrostatic forces \cite{derjaguin1941acta,verwey1947theory}. In water solution, the latter are characterized by the double layer potential and Debye-Hückel screening length ($\kappa^{-1}$). Later, Lifshits and colleagues demonstrated a deep intrinsic relation between Casimir and van der Waals forces \cite{lifshitz1992theory}. Historically, van der Waals interaction is usually referred to short-range, while Casimir to long-range separation distances, respectively. However, their physical origin is essentially the same \cite{lifshitz1992theory}. 

Casimir forces were successfully experimentally measured in several configurations using a torsion pendulum and microresonators \cite{lamoreaux1997demonstration,bressi2002measurement}. However, a similar Casimir problem may occur not only in a sterile vacuum environment, as was studied in these earlier experiments, but also in a crowded water solution, such as a suspension of chemically synthesized metallic nanoflakes \cite{chen2016rapid}. This brings us to the first step in formulation of the self-assembled polariton problem.

\subsection{Basic principle}

Fig. 1a illustrates the basic mechanisms enabling the self-assembled microcavities: a water solution of cetrimonium bromide ligand molecules (CTAB) hosts floating gold nanoflakes. When two flakes approach each other, two types of interaction emerge between them - the Casimir (van der Waals) interaction due to the vacuum fluctuations of the electromagnetic field in the cavity formed between the mirrors, and the electrostatic interaction due to the ion absorbance from the solution by the metallic surface and formation of the double electric layers. The former interaction between two metallic surfaces filled with a homogeneous isotropic medium is attractive, while the latter is repulsive. Therefore, the joined action of the two may result in the existence of stable equilibria.
Plot of the total ground energy of the system (see Methods) as a function of the ligand molecules density and the cavity thickness for a fixed surface charge density of 100 mC/m$^2$ (equivalent to $\approx 0.6$ electrons per nm\sq) shows an existence of stable configurations for a range of the ligand densities, Fig. 1b. Increasing the ligand concentration leads to screening of the electrostatic repulsion, causing the Casimir attraction to dominate the interaction and reducing the equilibrium distance to zero in the limit of high concentration. In the opposite limit of low density the electrostatic repulsion dominates the interaction, and the stable equilibrium at a finite distance disappears. 

\begin{figure*}
\includegraphics[width=1\textwidth]{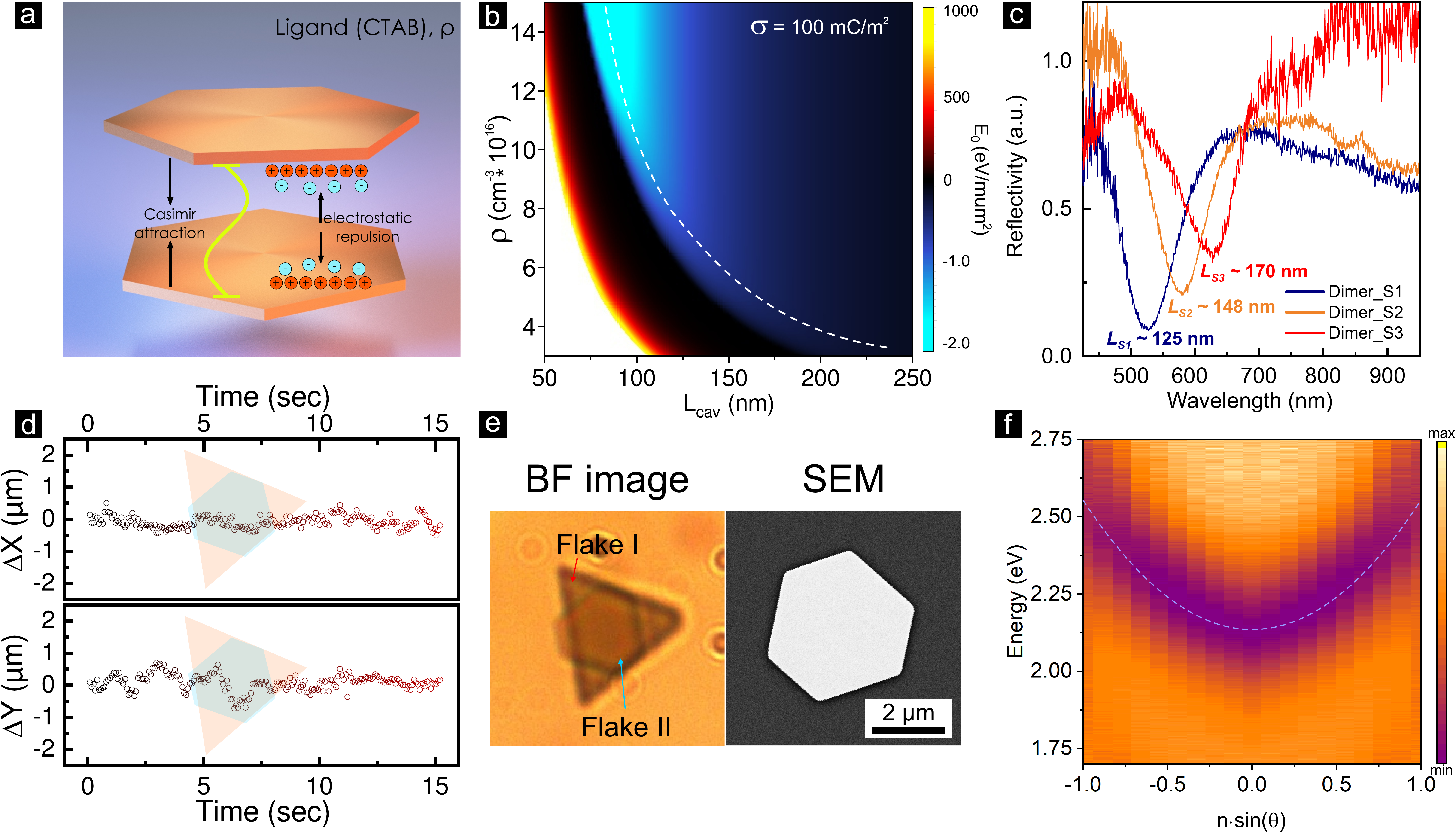}
\caption{\textbf{The system under study and the physical mechanism beyond its operation.} (a) Sketch of the system: two parallel gold flakes floating in a water solution of a ligand (CTAB). The two flakes are in a metastable equilibrium at a distance $L_{cav}$ from each other thanks to the joined action of the attractive Casimir force and the repulsive electrostatic double layer force. (b) The ground energy (sum of the Casimir and electrostatic double layer energies) of the system of two 30 nm thick gold flakes in a CTAB water solution ($n$=1.38) with the surface charge density $\sigma=$ 100 mC/m$^{2}$. The curve depicts evolution of the equilibrium distance with varying concentration. Note the inhomogeneous color scale. (c) Quasi-normal incidence reflectivity of the self-assembled cavities with $L_{cav} \approx$ 125 nm (blue), 148 nm (orange), and 170 nm (red), respectively. Different cavity thicknesses are obtained by diluting the original solution (blue curve), 1:1 (orange curve), and 1:2 (red curve) with deionized water, correspondingly. (d) Relative displacement of top and bottom nanoflakes within a dimer along $x$ and $y$ directions as a function of time. Note that while flakes do move independently, their relative displacement is always small in comparison to the lateral size of the flakes. This indicated the dimer stability not only in vertical, but also in lateral directions. (e) Bright-field and SEM images of an example gold nanoflake. (f) Angle-resolved reflection of an exemplary self-assembled microcavity with $L_{cav} \approx$ 148 nm, exhibiting a typical parabolic behavior.
}
\label{Fig1}
\end{figure*}

\subsection{Configuration I: Dimers}
To realize a stable Casimir microcavity experimentally, as shown schematically in Fig. 1a, we use chemically synthesized gold nanoflakes. Single crystalline gold nanoflakes (Fig. 1e), with average thicknesses of ~34$\pm$10 nm and a few microns in lateral dimensions, were synthesized using a rapid and seedless wet chemical method in aqueous solution (see Methods) \cite{chen2016rapid}. Freshly-prepared gold nanoflakes solution, containing CTAB ligand ($\sim$ 2 mM), was drop-casted onto a thin glass coverslip. The solution is encapsulated by another coverslip using a thin polydimethylsiloxane (PDMS) spacer, to prevent evaporation of the liquid.

After drop-casting, the non-aggregated nanoflakes slowly sediment towards the glass substrate, where they diffuse laterally due to Brownian motion until eventually colliding with another individual flakes or aggregates. In the simplest scenario, two isolated nanoflakes diffuse close to one another at a distance where the Casimir attraction becomes relevant, and form a stable nanoflake pair (see Fig. 1e and and supplementary movie S1.). This self-assembled dimer is also a \fp microcavity, whose equilibrium distance can be controlled by the concentration of ligand molecules in water solution, temperature, and light pressure. Once formed, such self-assembled microcavities remain stable for indefinitely long (in this study we have monitored the stability of self-assembled microcavities over the period of a few weeks without noticeable resonance degradation).

To access the quantitative information about the self-assembled microcavities, we performed reflection spectra measurements under quasi-normal incidence using an oil immersion 100$\times$ objective (Nikon, NA = 0.5), directed to a fiber-coupled spectrometer (see Methods). The measurements were performed on several gold nanoflakes microcavities, with different concentrations of CTAB ligands in the solution to control the equilibrium distance (Fig. 1c). To ensure reproducibility, we recorded reflectivity spectra from dozens of samples in the solution. The reflection data are highly reproducible (data available upon request).

Fig. 1c shows reflectivity spectra collected from three representative samples with different concentrations of CTAB in the solution. In order to tune the equilibrium distance, we have prepared three different gold nanoflake solutions by dissolving them in different amounts of de-ionized water. Dimer S1 was prepared from the original stock solution of the gold nanoflakes, whereas S2 and S3 were diluted with pure de-ionized water in 1:1 and 1:2 volume ratios, respectively. Dimers S1-S3 exhibit pronounced reflection dips at around $\sim$528, $\sim$578, and $\sim$626 nm, respectively (see Fig. 1c). These resonances correspond to the fundamental \fp cavity modes. The obtained reflectivity data allow us to extract the equilibrium distances using the transfer matrix method, and results in $L_{cav}$ of $\sim$125 nm, $\sim$148 nm, and $\sim$170 nm for Dimers S1, S2, and S3, correspondingly. Therefore, the equilibrium occurs at longer separation distances for more diluted samples, which is in qualitative agreement with theoretical plots in Fig. 1b. These observations clearly show that the cavity resonance can be tuned by varying the concentration of ligands in the solution (Fig. 1c).

To provide further insight into the physics of self-assembled microcavities, we performed angle-resolved reflectivity measurements using an inverted optical microscope equipped with an oil-immersion objective (100$\times$, NA = 1.3). The dispersion plot for reflectivity shown in Fig. 1f exhibits a characteristic of parabolic dispersion, thereby confirming the flat FP microcavity formation. More examples of dispersion plots are shown in supplementary Fig. S1). The obtained results indicate that the system is stable and can be used for photonic applications as a proper self-assembled microcavity.

To demonstrate that the stable self-assembled microcavity can be formed using this Casimir approach not only in the vertical direction, but also laterally, we followed the trajectories of the top and bottom nanoflakes within an exemplary microcavity over time (Fig. 1d). Due to Brownian motion, the microcavity as a whole is freely diffusing in the solution. However, the trajectories of the top and bottom flakes are not fully independent; instead, they are strongly correlated (see supplementary movie S2). This implies that the flakes within a dimer follow each other, but without forming a rigid inseparable aggregate. Noteworthy, this correlated movement occurs without a direct contact between the flakes, i.e. in a contact-free manner. Note that both nanoflakes within such a pair are positively charged at the surface by the CTAB ligand layers, and, thus, they should be electrostatically repelled. Yet, the flakes stick together for indefinitely long, which can only happen if the electrostatic repulsion is balanced by some attractive interaction. This attractive interaction is the Casimir force. However, this Casimir force acts not only in the vertical direction (as considered in Fig. 1b), but also laterally. This is an important finding of our work, as the existence and measurements of lateral Casimir forces is usually more challenging than vertical ones \cite{chen2002demonstration}. Numerical analysis of the lateral Casimir force goes beyond our current study, but the reader is referred to several useful theoretical approaches \cite{chen2002experimental,rodrigues2006lateral}.

\subsection{Configuration I: Trimers}
As was mentioned above, at the initial stage of the experiment, individual flakes freely diffuse in solution, until they find each other. Above, we have considered the simplest scenario of the dimer formation when two individual nanoflakes meet and form a stable Casimir microcavity, but the process is not stopped at that point. At a later stage, such a dimer can collide with another individual flake and form a stable trimer, which is schematically shown in Fig. 2a. The probability of that process depends on the concentration of nanoflakes in the solution and inter-flake interactions strengths, which allows controlling the relative population of dimers, trimers, and higher-order aggregates, in a manner similar to previously reported self-limiting aggregation behavior of silver nanoparticle colloids \cite{meyer2006self}.    

\begin{figure*}
\includegraphics[width=1\textwidth]{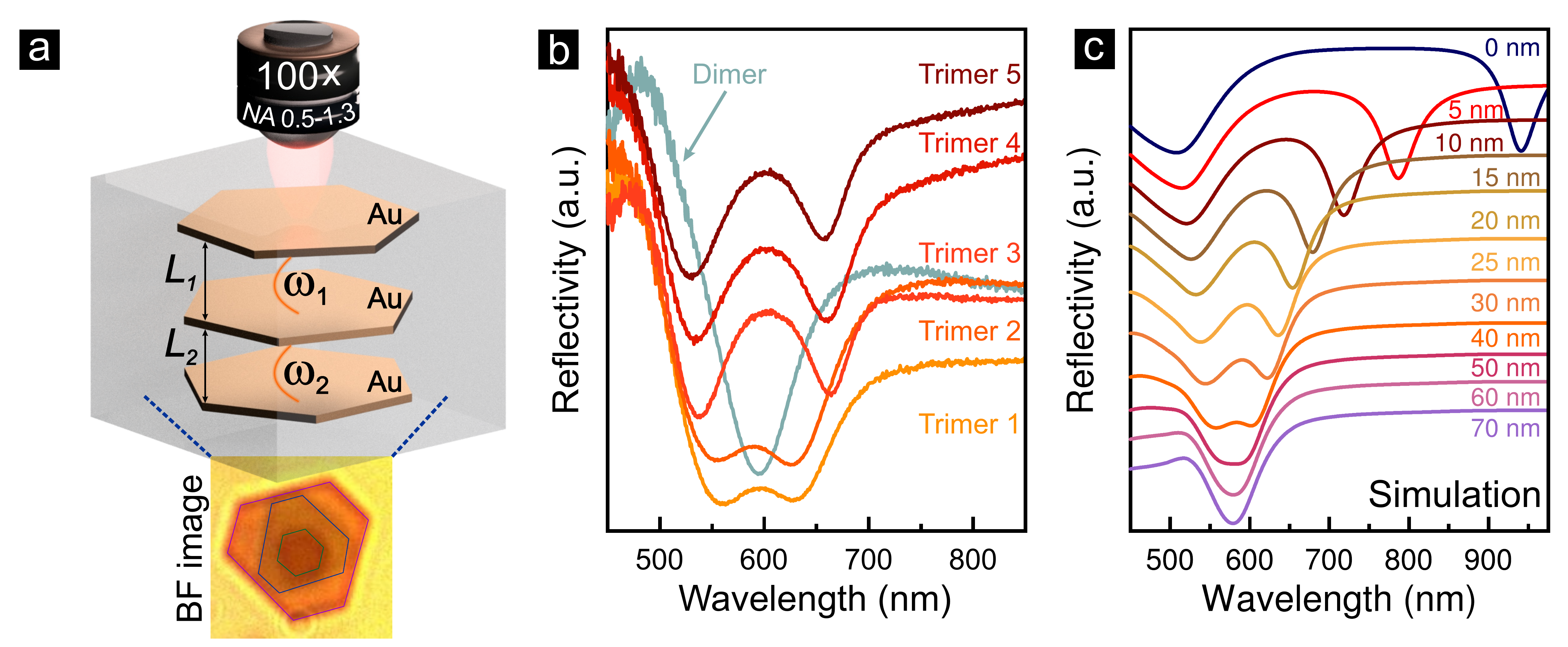}
\caption{\textbf{Self-assembled polaritons in trimers} 
(a) Sketch of the self-assembled trimer system. (b) Quasi-normal incidence reflectivity spectra for several self-assembled trimers with various geometrical characteristics. The grey-blue curve shows the reflectivity spectrum of a dimer for an easy comparison. (c) Calculated reflectivity spectra of three parallel gold flakes as a function of the thicknesses for middle-mirror (flake).
}

\label{Fig2}
\end{figure*} 

For a quantitative characterization we chose the simplest case of a multi-stack system, -- a trimer (higher-order aggregates are discussed in supplementary Fig. S2). We collected normal incidence reflectivity spectra from several trimers. The obtained results are summarized in Fig. 2b. It is noteworthy that samples were prepared from the diluted mixture solution of stock batch and de-ionized water with volume ratio of 1:1. Just like in the case of dimers, trimers form a stable aggregate both vertically and laterally (see an example trimer in supplementary movie S3). The corresponding trajectories of the flakes are shown in supplementary Fig. S3 and supplementary movie S4.

In contrast to dimers, trimers exhibit characteristic double-dip reflection spectra (see Fig. 2b-c). The origin of this double-dip behavior lies in the interaction of FP modes hosted by each cavity separated by the semi-transparent middle mirror. This situation is similar to the quantum mechanical problem of two potential wells separated by a penetrable barrier, where the solutions are symmetric and anti-symmetric combinations of the individual well's bound states \cite{landau2013quantum}.  
In the optical counterpart, the electromagnetic mode hosted by one of the cavities tunnels to the other half of the system at a rate determined by transmission coefficient of the separating mirror. This tunneling forms symmetric and anti-symmetric hybrid eigenmodes with the splitting between the two determined by the tunneling rate $w \sim e^{-k'' L}$ with $k''$ being the imaginary part of the wave vector inside the mirror material. This view of the problem is similar to the symmetric and anti-symmetric polariton branches in the Rabi-like problem \cite{rabi1937space}. Following these similarities, such three-mirror geometries were recently interpreted in terms of hybrid light-matter polaritonic states \cite{junginger2020tunable,berkhout2020strong,baranov2020ultrastrong}.

To illustrate this mechanism, we analytically calculated normal incidence reflection spectra of the trimer with the middle mirror thickness ranging from 70 nm down to 0 nm , corresponding to completely intransparent middle mirror and the absence of the middle mirror, respectively, Fig. 2c.
For thick middle mirror the tunneling rate vanishes and the reflection spectrum exhibits a single dip associated with the \fp eigenmode of the top cavity excited by the incident wave.
Reducing the middle mirror thickness allows tunneling and opens the splitting between two hybrid modes.
Such reduction of thickness leads to monotonic increase in the reflection dip splitting (Fig. 2c), until the lower-energy branch approaches the 1st order, while the higher-energy branch, correspondingly, the 2nd order FP mode of the cavity formed by top and bottom mirrors. This limiting case represents the maximum splitting possible for the three-mirror configuration.

Experimentally, we observe several different splittings (Fig. 2b), which indicates different thicknesses of the middle mirror. By fitting the experimental reflectivity curves using the transfer matrix method, we extracted parameters, including the thicknesses of the middle mirrors and apparent Rabi splittings. The results are summarized in supplementary Table I. It is worth mentioning that the initial values for cavity thicknesses between the top-middle and middle-bottom configurations were fixed to be identical.  The thicknesses for outer mirrors (top and bottom) were set to 25 nm. To perform fitting for the data presented in Fig. 2b, the thickness of the middle mirror was varied. Theoretically calculated fitting curves are shown in supplementary Fig. S4-S5. The extracted thicknesses of the middle mirrors in these experiments agree well with the atomic force microscopy (AFM) data of typical nanoflakes (see supplementary Fig. S6). Rabi splitting in the trimers increases when thickness of the middle mirror decreases (see supplementary Table I), as expected from theoretical analysis in Fig. 2c.

\subsection{Configuration II}
After discussing self-assembled cavities consisting of two or three nanoflakes in water solution, we switch our attention to an alternative cavity configuration, consisting of only one nanoflake in solution and a thin thermally-evaporated gold film used as another mirror (Fig. 3a). Such configuration has a number of advantages, namely a rapid formation of microcavities, and a possibility of using a spacer layer, which makes this option more versatile. Just like configuration I, this approach allows forming a self-assembled microcavity with an equilibrium distance between the film and the flake given by electrostatic repulsion and Casimir attraction.

\begin{figure*}[hbt!]
\includegraphics[width=1\textwidth]{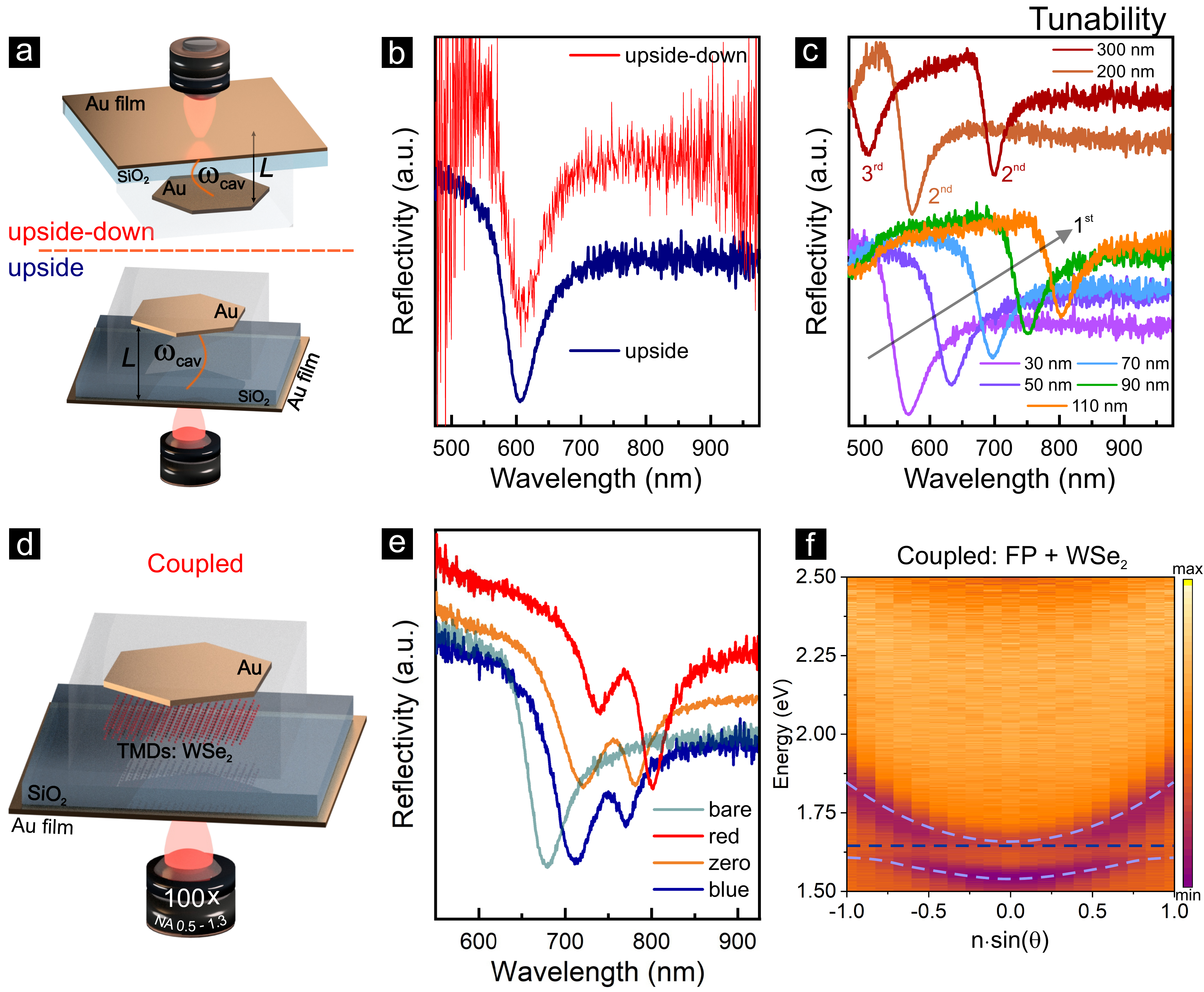}
\caption{\textbf{Microcavity consisting of a floating gold nanoflake and a planar gold mirror and formation of polaritons.} 
(a) Sketch of a self-assembled Fabry-Pérot microcavity in Configuration II. (b) Quasi-normal incidence reflectivity spectra for the microcavities with upside (blue) and upside-down (red) configurations, respectively. (c) Tuning the resonance of microcavities with various thicknesses of SiO$_2$ spacers. (d) Sketch of a self-assembled microcavity coupled to few layers of WSe$_2$. 
(e) Quasi-normal incidence reflectivity of the cavity coupled to various thicknesses of few layer WSe$_2$, showing pronounced Rabi splitting and correspondingly microcavity-exciton polariton formation. Note that red, orange, and blue curves show various detunings of cavity resonance with respect to the exciton resonance of WSe$_2$. (f) Dispersion measurement for the polaritonic microcavity loaded with few layer WSe$_2$, exhibiting a pronounced mode anti-crossing. Dashed lines are guides for the eye.
}

\label{Fig3}
\end{figure*}

Using this configuration we firstly check the role of gravity in the formation of stable Casimir microcavities. Fig. 3b shows reflection spectra of a single cavity collected in an upside and upside-down configurations. The reflectivity spectra in both cases exhibit dips at around 600 nm, suggesting the equilibrium distances in both cases are the same. This suggests that gravity plays a minor role in formation of these microcavities and the equilibrium distance is determined by the interplay of the Casimir attraction and the electrostatic repulsion.

Configuration II allows controlling the thickness of the metal film (which affects the $Q$-factor, see supplementary Fig. S7) and the thickness of the SiO$_2$ on top of the film (which affects the resonance frequency of the microcavity and allows its tunability, Fig. 3c). Noteworthy, the thickness of the spacer can be in principle significantly increased, so that the resulting FP cavity resonances will appear in the mid-infrared range, thus potentially expanding this self-assembled approach to strong coupling with molecular vibrations (VSC). In addition, Configuration II gives rise to a possibility to realize Casimir microcavities not only with gold, but also with silver nanoflakes (see Methods and supplementary Fig. S7).

Finally, by introducing an excitonic layer in between the film and a flake, configuration II allows realization of polaritonic states as mixtures of FP cavity photons and excitons in the excitonic material. This polariton is self-assembled and tunable. To realize such scenario experimentally, we choose WSe$_2$ motivated by its high oscillator strength and appropriate resonance of the A-exciton \cite{li2014measurement}. A few-layer WSe$_2$ was transferred on the SiO$_2$ spacer covering the gold film (see Methods), and subsequently covered by the nanoflake floating in water solution. Normal incidence reflectivity spectra measured from several such systems show pronounced Rabi splitting in the data (Fig. 3e). The corresponding dispersion measurement for the polaritonic microcavity loaded with a few-layer WSe$_2$, exhibits a pronounced mode anti-crossing, which proves the system is in the strong coupling regime.

\subsection{Active tuning}

The system shown in Fig. 4a represents a damped optomechanical resonator, with the restoring force being the joint Casimir-electrostatic potential, while the friction provided by the hydrodynamics of water. As any optomechanical system, this configuration allows tuning the cavity resonance actively by applying external stimuli. In this case we use modulated laser light to exert pressure on the nanoflake and, thus, displace it from its equilibrium (see Methods). This scenario is schematically shown in Fig. 4a.

\begin{figure*}[hbt!]
\includegraphics[width=1\textwidth]{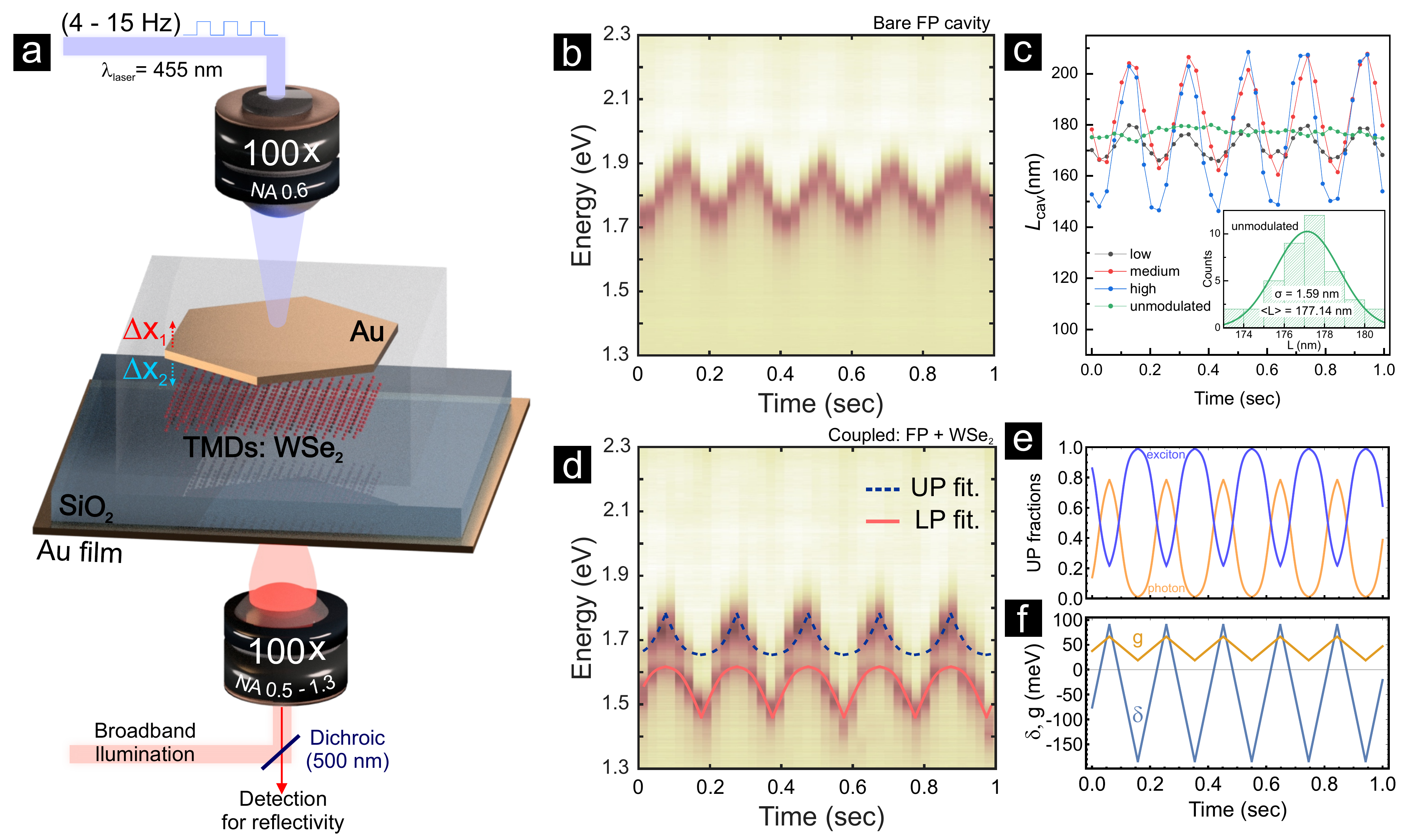}
\caption{\textbf{Actively tunable microcavities and polaritons} 
(a) Schematic drawing of the setup for active modulation of the microcavity by modulating the floating gold flake along vertical direction with an optical pressure - laser. Time-resolved (b) reflectivity spectra and (c) extracted total thicknesses for empty FP cavity under optical modulation at frequency of 5 Hz using laser irradiation with various powers of low (14.59 $J/cm^2$), medium (49.04 $J/cm^2$), and high (158.9 $J/cm^2$), respectively. The amplitude of oscillations depends both on the intensity and the frequency of modulated light. The closer the modulated frequency to the eigen frequency of the optomechanical oscillator (given by $\omega_{res}=\sqrt{k_{eff}/m_{eff}}$, where $k_{eff}$ is the effective spring constant and $m_{eff}$ is the effective mass of the oscillator). The displacement should be linear with the intensity of the drive in the small displacement limit. (d) Time-resolved reflectivity spectra of polariton system containing few layer of WSe$_2$ inside the FP cavity. Extracted (e) exciton-photon compositions for upper polariton, and (f) coupling strength (orange) and detuning (grey-blue), respectively.
}

\label{Fig4}
\end{figure*}

We start by modulating the empty FP cavity in configuration II. Fig. 4b shows the corresponding reflectivity profile, with the reflection dip oscillating in between $\sim$1.7 and 1.9 eV with time. This corresponds to vertical displacement of the nanoflake by $\sim \pm$ 20 nm with respect to the equilibrium position as the laser is exerted. The reflection as a function of time can be modelled by the transfer-matrix method (see Methods), which allows extracting the vertical displacement quantitatively (shown in Fig. 4c). It is important to mention that the modulation speed and depth depend on the parameters of the laser used in these experiments. Higher incident intensity can lead to larger vertical displacements. This is summarized in supplementary Fig. S8-S9. The modulation frequency was controlled by a mechanical chopper. Here the frequency range of 4-15 Hz was studied. The resonant frequency of the optomechanical system is given by the textbook expression $\omega_{res}=\sqrt{k_{eff}/m_{eff}}$, where $k_{eff}$ is the effective spring constant and $m_{eff}$ is the effective mass of the oscillator. Since this motion occurs in water, the effective mass is likely increased by the hydrodynamic drag and correspondingly the resonant frequency is small. This principally limits the modulation frequency and the amplitude of this optomechanical system.

Laser-induced modulation of the cavity resonance can be caused by one of the following phenomena: heating or light pressure. To differentiate between these scenarios, we performed additional experiments. In the first experiment, we illuminated the microcavity from below (the side of the static mirror). In this case, the system rapidly and irreversably lost its equilibrium vertical position. This behavior is in sharp contrast with the situation shown in Fig. 4b, where the microcavity was illuminated from top (the side of the movable mirror) and could be controllably modulated by light. These observations can be explained by the asymmetric nature of the combined electrostatic-Casimir potential near the equilibrium position (Fig. 1b). The heating scenario is thus unlikely, since it should not depend on the direction of laser light. The light pressure effect is, on the contrary, likely because it is consistent with the asymmetry of the electrostatic-Casimir potential.

In the second experiment, we recorded time-resolved reflectivity of the system in the absence of the modulating laser, showing Brownian fluctuations of the microcavity resonance. To visualize the heating effect, we turned the laser on and modulated the system at 5 Hz (see supplementary Fig. S10). The result showed a significant red-shift of 0.15 eV in the cavity resonance upon laser irradiation, in comparison to the Brownian regime. After a few seconds, the system returned to the equilibrium and showed a stable modulation behavior. Thereafter the driving laser was switched off, and the system showed an immediate blue-shift of the resonance, which eventually equilibrated to the original Brownian regime after a few seconds of cooling. This relatively slow dynamics can be attributed to laser-induced heating and cooling effects and a related to them rearrangement of CTAB molecules, but the faster cavity modulation is attributed to light pressure, in agreement with the first control experiment descibed above.

To quantify the stability of the self-assembled microcavity in the unmodulated Brownian regime, in Fig. 4c we show the time-resolved evolution of the equilibrium distance and its average value $<L_{cav}>$ of around $\sim$ 177 nm. The standard deviation $\sigma_L$ is only about 1.6 nm, which is rather remarkable for a micron-size self-assembled system at room temperature and in water solution. The inset in Fig. 4c shows the distribution of the $L_{cav}$ extracted by the transfer matrix method.

After modulating the empty FP cavity, we turn to modulation of the polaritonic system containing a WSe$_2$ multilayer between the mirrors. 
Variation of the cavity thickness in this scenario 
has a two-fold effect: not only it enables modulation of the bare \fp mode energy, but it also modifies the vacuum field of the cavity thus affecting the cavity-exciton coupling strength and the wave-function of the polaritonic eigenstates.
Positions of the two reflection minima, corresponding to upper and lower polariton modes, vary in time as seen in Fig. 4d. The reflection minima red-shift by $\sim 0.2$ eV compared to the empty cavity, which is due to the high background refractive index ($\sim 4$) of the WSe$_2$ layer \cite{munkhbat2018self}.

The recorded time-resolved reflection map (additional examples can be found in supplementary Fig. S11) allows extracting parameters of the coupled system as it is modulated. To that end, we fit the spectral positions of reflection minima with the eigenvalues of a periodically modulated \JC Hamiltonian (see Methods). 
Fig. 4f compares the resulting coupling strength $g(t)$ with the cavity-exciton detuning $\delta(t)=\omega_{cav}(t)-\omega_{exc}(t)$.
The \JC description of any polaritonic system predicts that once the detuning significantly exceeds the coupling strength, $|\delta(t)|>g(t)$, the system can be tuned out of the strong coupling regime.
Such behavior is clearly visualized in time-dependent Hopfield coefficients of the system showing nearly 100$\%$ photonic or excitonic character of the eigenstates at times corresponding to the minima of $g(t)$. Conversely, at time instances near the maxima of $g(t)$, the eigenstates of the system are polaritonic with nearly equal fractions of the photonic and excitonic components, Fig. 4e.

\section{Conclusion}
To conclude, we have presented a platform for self-assembled optical cavities and polaritonic states enabled by the joined action of the attractive Casimir and repulsive electrostatic interactions. Importantly, the self-assembled and tunable microcavitites studied here, on the one hand, exhibit pronounced optical resonances in technologically relevant visible spectral range, and, on the other hand, are highly stable in both vertical and lateral directions. The standard deviation of the $L_{cav}$ is as small as 1.6 nm at room temperature, which is on the order of 1$\%$ of the equilibrium cavity thickness. The long-term cavity stability is also remarkable, as the resonances remain almost unchanged for a period of at least a few weeks.

Moreover, our platform enables not only ordinary \fp microcavities, but also vertical multi-mirror aggregates with complicated modal structure, as well as polaritonic states that are obtained by loading the resulting cavities with excitonic films (such as WSe$_2$ layers). By modulating the system with a chopped laser light, we can actively control the polaritonic eigenstates and tune the system in and out of the strong coupling regime.
These findings open possibilities for exploring self-assembled Casimir microcavitites as sensitive and tunable polaritonic platforms in opto-mechanics \cite{eichenfield2009picogram}, nanomachinery \cite{zhao2019stable}, polaritonic chemistry \cite{Angew16}, and other promising cavity-induced applications. The possibility to actively tune the cavity in and out of strong coupling regime is especially important in the latter case, as it allows direct comparison between strongly and weakly coupled scenarios on exactly the same samples.  

\vspace{5mm}
\noindent \textbf{Acknowledgments}
The authors acknowledge financial support from the Swedish Research Council (under VR Miljö project, grant No: 2016-06059 and VR project grant No: 2017-04545), Knut and Alice Wallenberg Foundation, and Chalmers Excellence Initiative Nano.

\section{Methods}

\subsection{Casimir and DLVO calculations:}
Casimir (van der Waals) potential of two gold mirrors in a solution was calculated using the Lifshitz framework \cite{lifshitz1992theory}. The potential per unit area is obtained by the integration over the imaginary frequency $\omega=i\xi$ (with the parameter $\xi$ acquiring real values):
\begin{equation}
    U_{vdw}=\frac{\hbar}{2\pi}\int_0^{\infty} {d\xi \int {\frac{d^2 \mathbf{k_{||}} }{(2\pi)^2} \ln \det \mathbf{G} } }
\end{equation}
where $\mathbf{k_{||}}$ is the in-plane component of the wave vector in the gap region of thickness $L$, $\mathbf{G}=\mathbf{1}-\mathbf{R}_1 \times \mathbf{R}_2 e^{-2 K_0 L}$, and 
\begin{equation}
    \mathbf{R}_i=\begin{pmatrix}
r_i^{ss} & 0\\
0 & r_{i}^{pp}
\end{pmatrix}
\end{equation}
is the reflection operator for $i$-th side of the system ($i=1,2$); $r_{i}^{q}$ are the Fresnel reflection coefficients for $i$-th subsystem and polarization $q$ evaluated at the imaginary frequency. $K_0=\sqrt{\mathbf{k_{||}}^2 + \xi^2/c^2}$ is the z-component of the wave vector in the gap between the two mirrors evaluated at the imaginary frequency.

For calculations, we assumed the constant refractive index of the solution of $n=1.38$. The permittivity of gold at imaginary frequencies was evaluated by using the Drude model $\varepsilon_{Au}=8-\omega_{p}^2/\omega(\omega+i\gamma)$ with $\omega_{p}=8.6$ eV and $\gamma=70$ meV approximating the Johnson and Christy experimental data \cite{johnson1972optical}.

The electrostatic potential per unit area of two charged mirrors separated by the distance $L$ in a solution with relative permittivity $\varepsilon$ was estimated according to DLVO theory \cite{derjaguin1941acta,verwey1947theory}:
\begin{equation}
    U_{DLVO}=\frac{2\sigma^2}{\varepsilon_0 \varepsilon \kappa} e^{-\kappa L}
\end{equation}
where $\kappa=\sqrt{\frac{\rho q_0^2 z^2}{\varepsilon \varepsilon_0 k_B T}}$ is the inverse Debye-Hückel length, $\rho$ is the density of ions with valence $z$, and $\sigma$ is the surface charge density of the plates.

\subsection{Analysis of modulated cavities:}
Empty modulated cavities were analyzed by the standard transfer-matrix method. The reflection spectra at each time instance were fitted by reflection coefficient at normal incidence calculated assuming 30 nm thick gold mirrors, 55 nm SiO$_2$ spacer with $n=1.45$, and the solution refractive index of $n=1.38$ consistent with other calculations.

Reflection spectra from the modulated coupled structures were analyzed with the Hamiltonian approach. Time-dependent positions of reflection dips were fitted with eigenvalues $\omega_\pm$ of the \JC Hamiltonian $\omega_{\pm}= (\omega_{cav} + \omega_0)/2 - i(\gamma_{cav}+\gamma_0)/4 \pm \sqrt{g^2 - (\omega_{cav}-\omega_0+i(\gamma_{cav}-\gamma_0)/2)^2/4}$,
where $\omega_{cav} - i \gamma_{cav}/2$ is the cavity complex eigenenergy, $\omega_0 - i \gamma_0/2$ is the exciton complex energy, $g$ is the coupling strength. The cavity energy and the coupling strength were assumed to be modulated periodically with the triangular-like function (as suggested by the bare cavity reflection spectra dynamics):
\begin{align}
    \omega_{cav}(t)=\omega_{cav,0}+\delta \omega_{cav} f(\Omega t + \phi) \\
    g(t)=g_0 + \delta g f(\Omega t + \phi)
\end{align}
where $\Omega$ is the modulation frequency and $f(t)$ is the triangle wave:
\begin{equation}
    f(t)=4|t - \left \lfloor{t+1/2}\right \rfloor |-1.
\end{equation}
The cavity energy and the coupling strength modulation phases $\phi$ were assumed to be equal reflecting the fact that reducing cavity thickness increases the resonant energy, and increases the coupling constant at the same time.

The photonic and excitonic fractions of the polaritonic eigenstates are obtained as the corresponding elements of the Jaynes-Cummings Hamiltonian eigenvectors.

\bibliographystyle{naturemag}
\bibliography{Casimir}
\bibliographystyle{naturemag}

\end{document}